\documentclass[%
reprint,
 amsmath,amssymb,
 aps,
]{revtex4-2}

\usepackage{graphicx}
\usepackage{dcolumn}
\usepackage{bm}
\usepackage{hyperref}
\hypersetup{
colorlinks = true,	
urlcolor = blue,	
linkcolor= blue,	
citecolor = blue 
}
\usepackage{xcolor}

\begin{document}


\title{Periodic modulation of the space-filling nature of the turbulent flame leads to spiky heat release oscillations}

\author{Sivakumar Sudarsanan}

\author{Manikandan Raghunathan}
\author{Shwetha Viswesh}
\author{R. I. Sujith}%
\email{sujith@iitm.ac.in}
\affiliation{%
 Department of Aerospace Engineering, Indian Institute of Technology Madras, Chennai  600036, India
}%
\affiliation{Centre of Excellence for Studying Critical Transitions in Complex Systems, Indian Institute of Technology Madras, Chennai  600036, India} 





\date{\today}

\begin{abstract}
Spiky oscillations are characterized by slow-fast dynamics and are observed in excitable media such as neuronal membranes and cardiac cells. In a turbulent reactive flow system, we observe that the heat release rate exhibits self-sustained periodic, spiky oscillations in synchrony with the sinusoidal periodic acoustic pressure oscillations. 
These self-sustained oscillations are a consequence of thermoacoustic instability, which arises due to a positive feedback between the acoustic and the heat release rate fields.
One of the primary mechanisms for fluctuations in the heat release rate is the modulation in the topology of the flame, a thin interface separating the reactants and products. In this work, we explore the dynamics of the space-filling nature of the flame, quantified by its fractal dimension in relation to the spiky heat release rate oscillations in a turbulent reactive flow system. We discover that the periodic oscillatory dynamics in the space-filling nature of the flame lead to periodic, spiky heat release rate oscillations. Based on this result, we show that the spiky oscillations in the heat release rate can be approximated as $e^{\mathrm{sin}(\omega t)}$ during the dynamical state of generalized synchronization between the heat release rate and the acoustic pressure oscillations in a turbulent reactive flow system. In synchronization theory, generalized synchronization is characterized by an emergent functional relationship, $\Phi$, between the interacting subsystems. We unravel this emergent functional relation between the heat release rate and the acoustic pressure oscillations in a turbulent reactive flow system. It is intriguing that the dynamics of a far-from-equilibrium complex system can be represented by such simple mathematical relations.

\end{abstract}

\maketitle


\section{\label{sec:level1}Introduction}

\begin{figure*}  
  \centerline{\includegraphics[width=0.9\textwidth]{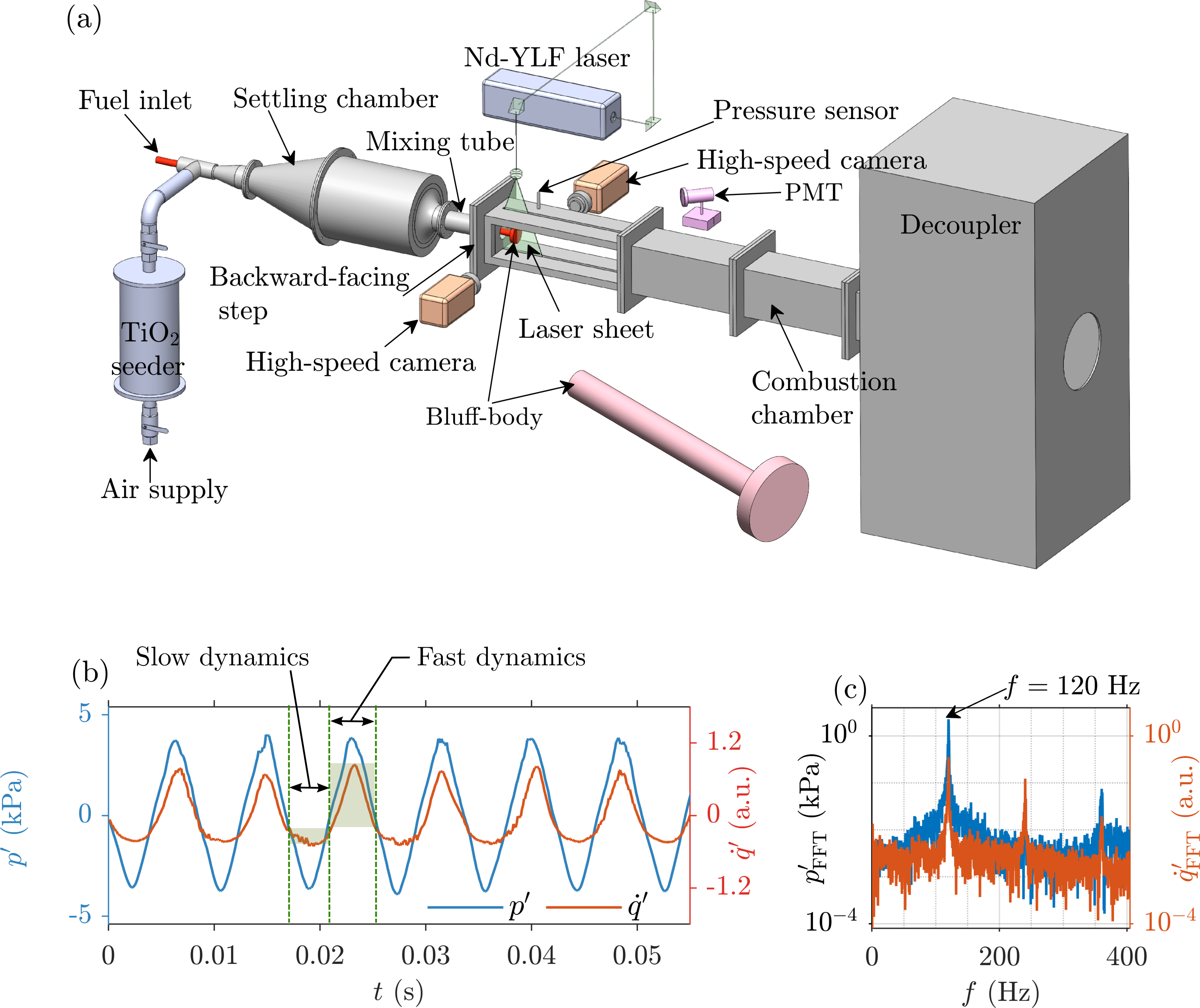}}
  \caption{\textbf{Experimental observation of periodic spiky oscillations in heat release rate.} (a) Schematic of the turbulent reactive flow system comprising air and fuel inlets,  a mixing tube, a combustion chamber, and a decoupler. The acoustic pressure, $p^\prime (t)$, is measured using a piezoelectric transducer mounted on the combustion chamber. A photomultiplier tube (PMT) outfitted with a $\mathrm{CH}^\star$ filter is used to obtain the global heat release rate, $\dot{q}^\prime(t)$. The airflow is seeded with TiO\textsubscript{2} particles, and an Nd:YLF laser is used as the light source to illuminate the flow field. The spatially resolved Mie-scattering images of TiO\textsubscript{2} particles and the CH$^{\star}$ chemiluminescence images are recorded using two different high-speed cameras. (b) The time series of the acoustic pressure (blue) and the heat release rate (red) oscillations at a Reynolds number value of $ 3.71 \times 10^4$. The acoustic pressure follows a sinusoidal variation. On the other hand, the heat release rate appears to be spiky, a flat variation followed by a sudden rise in its amplitude. (c) Amplitude spectrum of the acoustic pressure and the heat release rate oscillations. Both $p^\prime$ and $\dot{q}^\prime$ oscillate with the same frequency of 120 Hz. }
\label{fig: experimental setup}
\end{figure*}

Spiky oscillations are characterized by a slow variation followed by a rapid rise in the amplitude of a variable, and are observed in the membrane potential of neuronal membranes \citep{hodgkin1952quantitative, izhikevich2007dynamical} and cardiac cells \citep{noble1962modification}. In a turbulent reactive flow system, we observe that the heat release rate exhibits self-sustained periodic, spiky oscillations in synchrony with the sinusoidal acoustic pressure oscillations (Fig.~\ref{fig: experimental setup}) \cite{kasthuri2020recurrence, pawar2017thermoacoustic}. One of the primary mechanisms responsible for the fluctuations in the heat release rate is the modulation in the topology of the flame, a thin interface separating the reactants and products \citep{candel2002combustion,renard2000dynamics}. 
The topology of a turbulent flame can be characterized by its fractal dimension, which quantifies the space-filling nature of the turbulent flame, i.e., how the topology of the flame, comprising multiple scales of corrugations, fills the spatial domain. During the self-sustained periodic oscillatory state, the fractal dimension of the flame oscillates periodically \citep{raghunathan2020multifractal}. In this work, we study the spiky oscillations in heat release rate in relation to the dynamics of the space-filling nature of the flame in a turbulent reactive flow system. 

The turbulent reactive flow system consists of the flame, the acoustic field, and the hydrodynamic field interacting in a nonlinear manner \citep{candel2002combustion, lieuwen2005combustion, sujith2020complex}. A positive feedback loop arising from the nonlinear interaction between the subsystems leads to a change in the dynamical state of acoustic pressure from chaotic to self-sustained periodic oscillations \citep{nair2013loss,sujith2020complex}. 

The interaction between flame and vortex leads to wrinkles on the flame surface \citep{candel2002combustion, renard2000dynamics}. The vortices represent a connected large-scale region in the flow field with coherent vorticity \citep{hussain1983coherent}. Vorticity is a measure of the local rotation of the fluid parcel \citep{batchelor2000introduction,kundu2024fluid}. In backward-facing step combustors, as the vortex convects downstream along the shear layer, the reactants are entrained into the hot product side,
and distort the flame front, thereby increasing the surface area of the flame \citep{renard2000dynamics}. Furthermore,  small-scale vortices along the shear layer interact and self-organize at a large scale \citep{schadow1992combustion,george2018pattern}. Consequently, the flame surface exhibits multiple scales of wrinkles, resulting in the sudden release of heat energy \citep{raghunathan2020multifractal, schadow1992combustion}.  

The localized intense heat release resulting from the interaction between the flame and vortices is referred to as a heat release rate pulse \citep{poinsot1987vortex, ducruix2003combustion}.  Based on the experimental observations, Matveev \textit{et al.}~\cite{matveev2003model} modeled the intense heat release rate oscillations as a kicked oscillator by assuming localized vortex burning in space and time. When the heat release rate fluctuations are in phase with acoustic pressure fluctuations, energy is added to the acoustic field; a necessary condition for the emergence of self-sustained oscillations and the occurrence of thermoacoustic instability \citep{rayleigh1878explanation, poinsot2005theoretical}. Thermoacoustic instability is often encountered in gas-turbine and rocket engines, characterized by ruinously large-amplitude oscillations in the acoustic pressure, which are undesirable due to the potential structural damage and operational safety of these engines \citep{candel2002combustion,lieuwen2005combustion, zinn2005combustion}.

The time series of global heat release rate oscillations, $\dot{q}^\prime(t)$, and time series of acoustic pressure oscillations, $p^\prime (t)$ during the sustained periodic oscillatory state at a Reynolds number of $ 3.71 \times 10^4$ are shown in Fig.~\ref{fig: experimental setup}b. Considering the variation of the amplitude of the signals shown in Fig.~\ref{fig: experimental setup}b, the acoustic pressure appears to be sinusoidal \cite{pawar2017thermoacoustic}.
On the other hand, the heat release rate oscillations exhibit a spiky variation in their amplitude. Specifically, this spiky variation is characterized by a relatively slow variation followed by a rapid rise in the amplitude. The epochs of this slow-fast dynamics are highlighted in Fig.~\ref{fig: experimental setup}b.

Recently, Kasthuri \textit{et al.}~\cite{kasthuri2020recurrence} revealed the presence of slow and fast dynamics in the heat release rate using recurrence analysis. Such a slow-fast time dynamics is a characteristic feature of a spiking phenomenon in excitable media \citep{hodgkin1952quantitative, izhikevich2007dynamical, noble1962modification}. Excitable media in a stable resting state can be triggered by a sufficiently strong perturbation, leading to the propagation of a wave or pattern \citep{zykov2018spiral}. A characteristic feature of excitable media is its refractory time period. It refers to the duration in which the medium returns to its resting state after an episode of activation, before it can allow the propagation of a subsequent wave \citep{dawes1956refractory, sinha2014patterns, moosavi2017refractory}. The refractory time period is associated with the slow-regulating process in the medium, and its interaction with a fast excitable subsystem gives rise to the separation of timescales or slow-fast dynamics \citep{strogatz2018nonlinear}.


A visual inspection reveals that $p^\prime$  and $\dot{q}^\prime$ exhibit synchronized dynamics during the sustained periodic oscillatory state (Fig.~\ref{fig: experimental setup}b). $p^\prime$ and  $\dot{q}^\prime$  oscillate with the same frequency of 120 Hz, corresponding to the fundamental, longitudinal acoustic mode of the combustion chamber (Fig.~\ref{fig: experimental setup}c). Pawar \textit{et al}. \citep{pawar2017thermoacoustic}  have studied the transition from chaotic to periodic acoustic pressure oscillatory state using the framework of synchronization theory. 

Synchronization is a key feature of a complex system comprising several interacting subsystems, where the dynamical behavior of different subsystems aligns in time \citep{pikovsky2001synchronization, manrubia2004emergence}. Based on the relationship among the dynamics of the subsystems, synchronization can be classified into several different categories, i.e., {complete synchronization, partial synchronization, phase synchronization, and generalized synchronization} \citep{pikovsky2001synchronization}. During generalized synchronization (GS), the dynamics of the system reduces to a low-dimensional manifold that represents the emergent functional relation between the dynamical states of the interacting subsystems \citep{rulkov1995generalized,abarbanel1996generalized, kocarev1996generalized, zheng2002transitions}. This functional relation, known as the generalized synchronization manifold, can be smooth, smooth-nondifferentiable, or nonsmooth-fractal in nature \citep{paoli1989long, hunt1997differentiable, stark1997invariant}. 

As the turbulent reactive flow system transitions from chaotic to periodic oscillatory state, the synchronous dynamics between the heat release rate and the acoustic pressure exhibits a transition from a state of desynchronization to a state of phase synchronization, and eventually to a state of generalized synchronization \citep{pawar2017thermoacoustic}. Using curve fitting, Pawar \textit{et al.}~\citep{pawar2017thermoacoustic}  proposed that the spiky oscillations in heat release rate follow a functional form, $\dot{Q}^\prime(t) \sim \mathrm{sin}(\mathrm{sin}(\omega t))$, during the dynamical state of generalized synchronization. However, they did not explain the underlying physical mechanism responsible for the spiky oscillations in heat release rate. 



In this work, we study physical processes associated with the spiky heat release rate oscillations, specifically, the dynamics of the flame surface topology and its space-filling nature in a turbulent reactive flow system. Based on the insights from the physical processes, we study the emergent functional relation associated with the spiky heat release rate oscillations during the dynamical state of generalized synchronization. Our study provides the first evidence that the periodic dynamics in the fractal dimension, i.e., space-filling nature of the flame surface result in periodic, spiky oscillations in the heat release rate.

\section{Experimental setup}
The schematic of the experimental system, a turbulent combustor with a backward-facing step, is shown in Fig.~\ref{fig: experimental setup}a. The experimental setup comprises a settling chamber, a mixing tube, a combustion chamber, a bluff body, and a decoupler. The oxidizer (atmospheric air) is supplied to the settling chamber, which is used to dampen the fluctuations of the incoming air flow. We use liquefied petroleum gas (LPG) with a composition of $60 \%$  C\textsubscript{4}H\textsubscript{10} and $40 \%$  C\textsubscript{3}H\textsubscript{8} as the fuel. The fuel is supplied to the mixing tube through a hollow shaft of diameter 16 mm, supported by a rack and pinion mechanism attached to the settling chamber. The hollow shaft has four circumferentially arranged holes, each with a diameter of 1.7 mm, through which the fuel is injected into the mixing tube, where the air from the settling chamber mixes with the fuel. The fuel-air mixture enters the combustion chamber through a backward facing step. 
The combustion chamber has a cross-section of $90 \times 90$ $\textrm{mm}^2$ and a length of  1100 mm. Two quartz glass windows of $90 \times 360$ $\textrm{mm}^2$ are provided on both the side walls of the combustion chamber for optical access. We ignite the reactant mixture using a spark plug to start the combustion experiment. A bluff body is used for flame stabilization. The bluff body is fixed at a location 30 mm downstream of the backward-facing step of the combustion chamber with the help of a central shaft attached to the settling chamber. The bluff body is a circular disk with a diameter of 55 mm and a thickness of 10 mm.

The outlet of the combustion chamber is connected to a chamber whose cross-sectional area is much larger than that of the combustion chamber. This large chamber is referred to as a decoupler, which has dimensions of $1000 \times 500 \times 500$ $\textrm{mm}^3$. The decoupler isolates the system from ambient fluctuations and maintains the acoustic boundary condition, $p^\prime_{\mathrm{outlet}}=0$, at the outlet of the combustion chamber.   

The mass flow rates of fuel and air are independently controlled by mass flow controllers (Alicat Scientific MCR series) with an uncertainty of $\pm 0.8\%$  of measured reading $\pm 0.2 \%$ of full-scale reading.  Reynolds number is defined as $\mathrm{Re}=\bar{u} d/ \nu_\mathrm{reactants}$, where $\bar{u}$ is the average velocity of the fuel-air mixture entering the combustion chamber, $d$ is the diameter of the bluff body, and $\nu_\mathrm{reactants}$  is the kinematic viscosity of the mixture.  Kinematic viscosity is calculated by considering the composition of the mixture \citep{bird1961transport}. We observe a dynamical state of generalized synchronization corresponding to the operating conditions with a fuel flow rate of $\dot{m}_\mathrm{fuel}=0.92$ g/s, and an air flow rate of $\dot{m}_\mathrm{air}=16.32$ g/s. The corresponding Reynolds number is $3.71 \times 10^4$. {The uncertainty in the Reynolds number is $\pm{493}$, calculated based on the uncertainty of the mass flow controllers.}

A piezoelectric pressure transducer (PCB 103B02) is used to measure the acoustic pressure fluctuations, $p^{\prime}(t)$, inside the combustion chamber. The pressure transducer is mounted to the wall of the combustion chamber 120 mm downstream of the backward-facing step using a T-joint mount. A 10 m long, 4 mm inner diameter semi-infinite waveguide is connected to the transducer mount to reduce the frequency response of the probe. 
The sensitivity of the pressure transducer is 223.4 mV/kPa, with an uncertainty of $\pm 0.15$ Pa. The signals from the piezoelectric pressure transducer were recorded using a data acquisition system (NI DAQ-6346, 16 bit). 

The chemiluminescence intensity of the $\mathrm{CH}^\star$ radical in a flame is a measure of the heat release rate \citep{hardalupas2004local, ikeda2000measurement}.  A photomultiplier tube (PMT; Hamamatsu H10722-01) outfitted with a $\mathrm{CH}^\star$ band-pass filter (a narrow band filter of peak at 435 nm with 10 nm full width at half maximum) is used to acquire the global heat release rate ($\dot{q}_{\mathrm{}}^{\prime}(t)$) of the flame. The heat release rate and the acoustic pressure signals are simultaneously acquired using the data acquisition system (NI DAQ-6346). The spatially resolved, local heat release rate fluctuations ($\dot{q}_{\mathrm{}}^{\prime}(\mathbf{x}, t)$) are acquired using the high-speed Phantom v12.1 camera,  outfitted with a $\mathrm{CH}^\star$ filter along with a 100 mm Carl-Zeiss lens. The chemiluminescence images are captured simultaneously with the acoustic pressure signal.

The Mie-scattering experimental method involves seeding the flow with tracer particles and imaging the illuminated flow field. A double-pulsed Nd:YLF laser with a single cavity (wavelength of 527 nm and a pulse width of 250 ns) is used as the light source for illuminating the flow field. The laser is operated in single-pulse mode at a repetition rate of 2000 Hz. A 2 mm thick laser sheet is created by expanding the laser beam using a spherical lens (600 mm focal length) and a plano-concave cylindrical lens (15 mm focal length). A rectangular quartz window (400 $\times$ 20 $\times$ 10 mm$^3$) is provided on the top wall of the combustion chamber. The laser sheet passes through this quartz window to illuminate the flow field. As the TiO\textsubscript{2} particles traverse the plane illuminated by the laser sheet, they scatter light, which is captured by a high-speed camera (Photron FASTCAM SA4) synchronized with the laser at 2000 fps. Thus, we obtain the spatial distribution of TiO\textsubscript{2} particles. The camera is equipped with a Carl-Zeiss 100 mm lens and outfitted with a narrow bandpass optical filter (527 $\pm$ 12 nm) to capture the light scattered by the seeding particles.

Titanium dioxide (TiO\textsubscript{2}) particles (Kronos, product 1071) with an average diameter of approximately 1~$\mu$m are used to seed the flow. For obtaining sufficient supply and uniform distribution of TiO\textsubscript{2} particles, a portion of the primary air stream is directed through a fluidized bed seeder, where the particles are thoroughly mixed with the incoming air. The air flow with uniformly distributed TiO\textsubscript{2} particles then reintroduced into the primary air flow upstream of the settling chamber. Corresponding to the Mie-scattering experiment, the flow rates are $\dot{m}_\mathrm{fuel}=1.09$ g/s and $\dot{m}_\mathrm{air}=16.72$ g/s, resulting in a Reynolds number value of $3.86 \times 10^4$. The uncertainty in the Reynolds number value obtained from the uncertainties of the mass flow controllers is $\pm506$.

\section{Spiky oscillation in the heat release rate is associated with the emergent large-scale structure \label{section: physical_mechanism_SpikyHRR}}

\begin{figure*}
    \centering
    \includegraphics[width=0.65\textwidth]{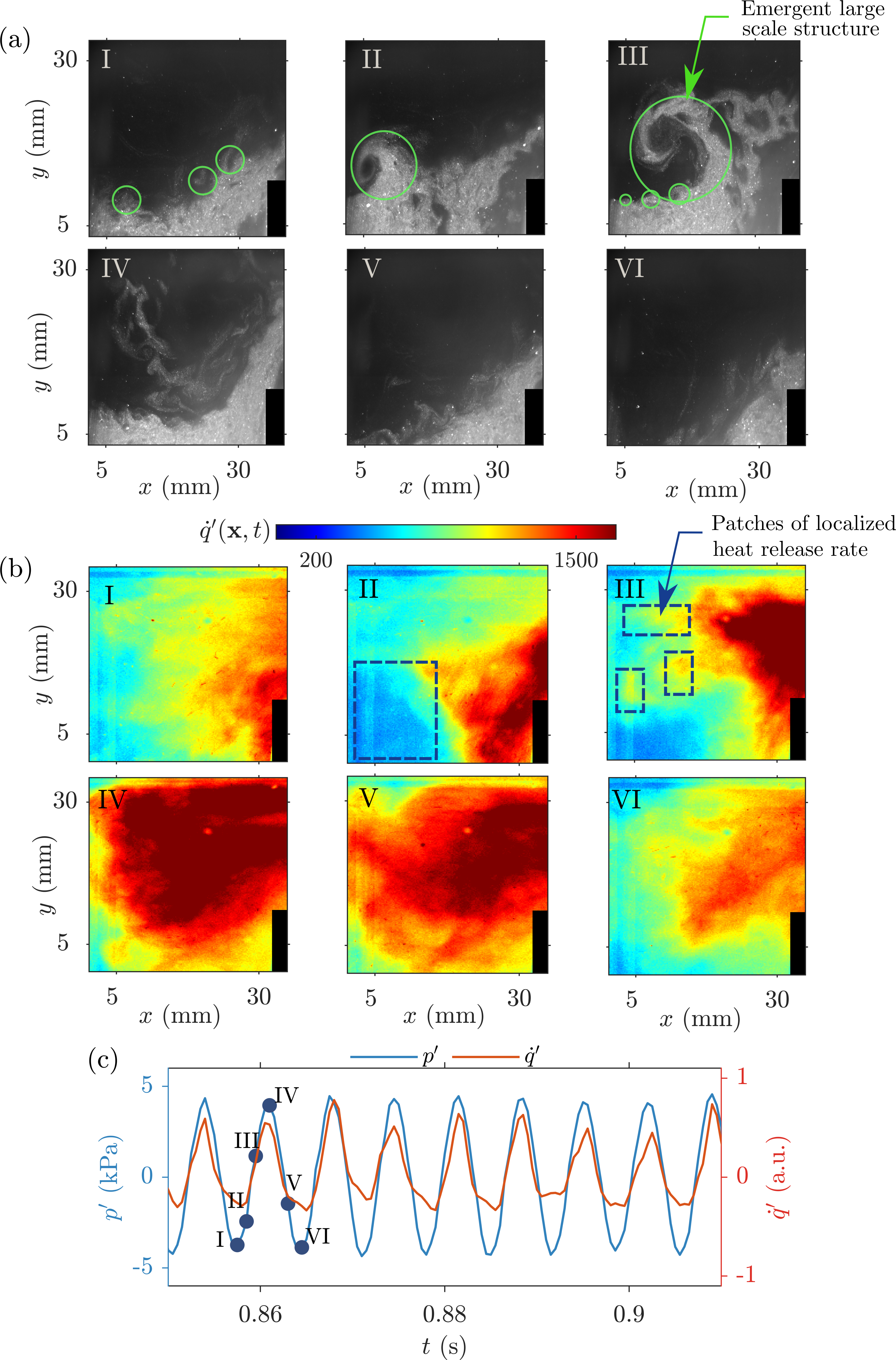}    
    \caption{\textbf{Sudden heat release arises due to the emergent large-scale structure.} (a) Planner Mie scattering images of TiO\textsubscript{2} particles and (b) chemiluminescence images of $\mathrm{CH}^\star$ radical corresponding to the instances c-I to c-VI. The chemiluminescence intensity is a qualitative measure of heat release rate. (c) The acoustic pressure overlaid with the global heat release rate oscillations.  Over one acoustic cycle, the sequence follows. (c-I) The flow field is characterized by small-scale structures indicated with circles in a-I. (c-II) Vortex roll-up entraining the fresh reactants into the hot product side, marked with an arrow in a-II. (c-III) The emergent large-scale structure, highlighted with circles in a-III. The localized patches of heat release rate are highlighted with rectangles in b-III. (c-IV, c-V) Sudden and high heat release rate over the spatial domain in b-IV. (c-VI) Eventually, the entrained reactants are burned, and the flow field is characterized by small-scale structures.}
    \label{fig: CH6 flameDynamics_physical}
\end{figure*}

The planar Mie scattering images of TiO\textsubscript{2} particles over one acoustic pressure cycle during the sustained periodic oscillatory state are shown in Fig.~\ref{fig: CH6 flameDynamics_physical}a. The time stamps corresponding to these images are shown in Fig.~\ref{fig: CH6 flameDynamics_physical}c. The regions with a high concentration of TiO\textsubscript{2} particles represent fresh reactants, whereas the regions with a low concentration of TiO\textsubscript{2} particles represent burned products \citep{stella2001three, pfadler2005measurement, steinberg2008measurements}.  

Figure \ref{fig: CH6 flameDynamics_physical}b shows the chemiluminescence images of $\mathrm{CH}^\star$ radical corresponding to the same instances as the Mie scattering images shown in Fig.~\ref{fig: CH6 flameDynamics_physical}a.  Chemiluminescence intensity of the $\mathrm{CH}^\star$ radical is a qualitative measure of heat release rate \citep{hardalupas2004local}. Figure \ref{fig: CH6 flameDynamics_physical}c presents the time series of acoustic pressure oscillations, $p^\prime (t)$, and the time series of global heat release rate oscillations, $\dot{q}^\prime(t)$. In this work, a variable with a superscript prime, such as $z^\prime$, denotes the mean-subtracted quantity,
defined as $z^\prime=z-  \langle z \rangle$, where $\langle \cdot\rangle$ represents the mean value.

Referring to Fig.~\ref{fig: CH6 flameDynamics_physical}a-I, at the minima of the acoustic pressure, the spatial distribution of reactants extends along the ${x}$ direction. At this instant, only small-scale structures (highlighted with circles in Fig.~\ref{fig: CH6 flameDynamics_physical}a-I), with a characteristic size of approximately 5 mm, are observed. Subsequently, a vortex roll-up, entraining the unburnt reactant mixture into the hot product side, is observed near the tip of the backward-facing step (highlighted with a circle in Fig.~\ref{fig: CH6 flameDynamics_physical}a-II). Consequently, the heat release rate is reduced near the inlet of the backward-facing step (highlighted with a rectangle) of the combustor (Fig.~\ref{fig: CH6 flameDynamics_physical}b-II). As the vortex, highlighted with an arrow mark in Fig.~\ref{fig: CH6 flameDynamics_physical}a-II, convects downstream, it grows, and the interaction between several small-scale vortices along the interface between the products and reactants leads to an emergent self-organized structure at a larger scale,  with a size of approximately 16 mm (Fig.~\ref{fig: CH6 flameDynamics_physical}a-III) \citep{george2018pattern}. The small-scale vortices and the emergent large-scale structure are highlighted with circles in Fig.~\ref{fig: CH6 flameDynamics_physical}a-III. Corresponding to this instant, localized patches of heat release rate begin to appear in the spatial domain (the localized patches are highlighted with rectangles in Fig.~\ref{fig: CH6 flameDynamics_physical}b-III). 

Followed by the emergence of large-scale structure, the area of the reactants (bright regions in the Mie-scattering images) reduced significantly (Fig.~\ref{fig: CH6 flameDynamics_physical}a-IV), and high chemiluminescence intensity values are observed over the spatial domain (Figs.~\ref{fig: CH6 flameDynamics_physical}b-IV, b-V). The emergent large-scale structure enhances the space-filling nature of the interface between the reactants and products, leading to a sudden and spiky heat release rate. At the end of the acoustic pressure cycle, the reactants entrained by the large-scale structure are burned, and the spatial distribution of reactants now extends along the ${x}$ direction (Fig.~\ref{fig: CH6 flameDynamics_physical}b-VI). The processes illustrated in Fig.~\ref{fig: CH6 flameDynamics_physical}a,b repeat for every acoustic cycle.

The emergence of the large-scale structure is associated with the slow convective timescale, i.e., vortices shed from the inlet step of the combustor entrain the reactants, convect downstream, and collectively interact to form a large-scale coherent structure (Fig.~\ref{fig: CH6 flameDynamics_physical}a-II,a-III). Subsequently, a sudden increase in the heat release rate on a fast timescale (Fig.~\ref{fig: CH6 flameDynamics_physical}b-IV).  Such slow-fast dynamics are a characteristic feature of spiky oscillations that are ubiquitous in excitable media \cite{hodgkin1952quantitative, izhikevich2007dynamical}.


The modulation in the topology of the flame surface is a major factor influencing the heat release rate dynamics \citep{steinberg2010flow}. Balachandran \textit{et al.}~\cite{balachandran2005experimental} showed that the global heat release measured with chemiluminescence and the estimation based on flame surface area were in close agreement in a turbulent premixed combustor subjected to inlet velocity perturbations. A significant velocity perturbation leads to roll up of the shear layer to form vortices, and the flame wraps around these vortices, followed by a significant increase in the heat release rate \citep{balachandran2005experimental}. Such interactions between the flame and the vortices, and the resultant modulations in the surface area of the flame, play a significant role in the emergence of thermoacoustic instability \citep{raghunathan2020multifractal,steinberg2010flow, venkataraman1999mechanism}. 

From the Mie scattering images of TiO\textsubscript{2} seeding particles, we identify the interface between regions of high and low particle number density. We refer to this interface as the flame contour, which represents the topology of the flame surface \citep{stella2001three, pfadler2005measurement, steinberg2008measurements}. Refer to Appendix \ref{app_A} for details on the methodology for identifying the flame contours. In the following section, we characterize the topology of the flame contours using the framework of fractal geometry \citep{mandelbrot1983fractal}.  

\section{Periodic oscillation of the fractal dimension of flame contours \label{section: physical_mechanism_SpikyHRR}}

Irregular geometries that exhibit structure across multiple scales are described using the concept of fractals, for example, coastlines \citep{mandelbrot1983fractal, mandelbrot1967long}, clouds \citep{lovejoy1982area}, river networks \citep{tarboton1988fractal}, electric discharge \citep{niemeyer1984fractal}, and the circulatory system of organisms \citep{west1999fourth}.  The variation of details, such as the length or area, with respect to the scale of observation of a fractal object is quantified by its fractal dimension. Unlike Euclidean geometric objects, fractal objects exhibit a non-integer, fractional dimension. For example, the coastline of Great Britain has a fractal dimension value of 1.25 \citep{mandelbrot1967long}. 

\begin{figure*}[ht]
    \centering
    \includegraphics[width=0.7\textwidth]{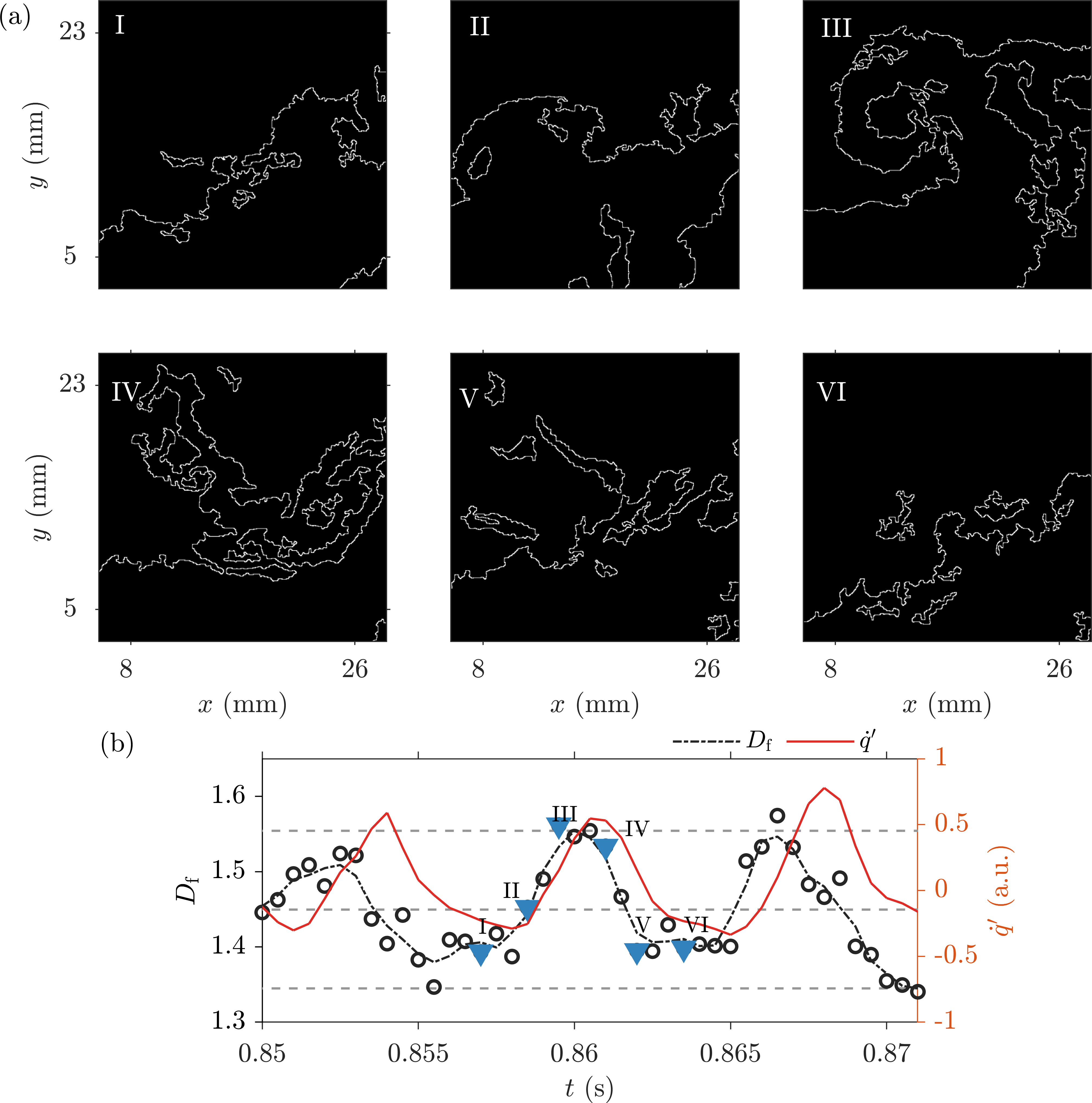}    
    \caption{ \textbf{The fractal dimension of the flame contours oscillates periodically during the sustained periodic oscillatory state.} (a-I to a-VI) Evolution of flame contours corresponding to the instances b-I to b-VI during the periodic oscillatory state. These flame contours are obtained from the Mie scattering images of TiO\textsubscript{2} particles shown in Fig.~\ref{fig: CH6 flameDynamics_physical}a. (b) The heat release rate oscillations overlaid with the corresponding fractal dimension values of the flame contours. At b-I, the flame contours extend along the ${x}$ direction. (b-II) Onset of vortex roll-up. (b-III) Emergent large-scale structure due to the vortex roll-up and interaction between small-scale vortices. (b-IV, b-V) Flame contours begin to disappear. (b-VI) Eventually, the flame contour extends along the ${x}$ direction, a similar structure to that at the beginning of the cycle. Corresponding to the instance, b-III, multiple scales of wrinkles enhance the space-filling of flame contours. The value of $D_\mathrm{f}$ oscillates periodically \citep{raghunathan2020multifractal}, suggesting periodic emergence of large coherent structures during self-sustained periodic acoustic pressure oscillations.}
    \label{fig: flameDynamics_spiky}
\end{figure*}


Fig.~\ref{fig: flameDynamics_spiky}a presents the dynamical evolution of the flame contours during an acoustic pressure cycle. The flame contour represents a two-dimensional cross-section of the flame surface, which is inherently a three-dimensional structure. We observe different scales of corrugations or wrinkles on the flame contours, a characteristic of the fractal topology of turbulent flames (Figs.~\ref{fig: flameDynamics_spiky}a-I to a-VI). These corrugations increase the total length of the flame contour, effectively increasing its space-filling nature. 

Considering the fractal topology of the flame contour, its total length can be expressed as a power law over a range of scales with fractal dimension as the scaling exponent \cite{mandelbrot1983fractal, Falconer_2003, sreenivasan1991fractals}, 
\begin{equation}
    L_{\mathrm{flame}} \sim \left( \frac{L_{\mathrm{o}}}{L_\mathrm{i}}\right) ^{D_\mathrm{f}-1}, 
\end{equation}
where  $L_{\mathrm{i}}$ and $L_{\mathrm{o}}$ are the inner and outer cut-off length scales \cite{cintosun2007flame, cintosun2007flame_conference, herbert2024comparison}. The inner cut-off scale corresponds to the smallest scales for the interaction of turbulent vortices with the flame surface. In turbulent reactive flows, the Kolmogorov length scale, $\eta$, is often chosen as an appropriate estimate for the inner cut-off scale \citep{gouldin1987application, gulder1991, gulder1995inner}. The outer cut-off scale represents the characteristic length scale of the flow, i.e., the step size of the backward-facing step combustor. The ratio $L_{\mathrm{o}}/L_{\mathrm{i}}$, depends on the Reynolds number of the flow \citep{gouldin1987application, thiesset2016geometrical}. 


The fractal dimension of the flame surface ($D_\mathrm{f}^\mathrm{3D}$) is related to the fractal dimension of the flame contours ($D_\mathrm{f}$) through a relation, $D_\mathrm{f}^\mathrm{3D}=D_\mathrm{f}+1$ \citep{mandelbrot1983fractal, chatakonda2013fractal}. Experimental studies have reported a fractal dimension value close to 7/3 for the premixed turbulent flame surface \citep{gouldin1987application, kerstein1988fractal}. 
   
The global heat release rate, $\dot{q}$, is proportional to the total length of the flame contours \citep{poinsot2005theoretical, peters2000turbulent}, therefore,
\begin{align}
    \dot{q} &\sim  \left(  \frac{L_{\mathrm{o}}}{L_\mathrm{i}}\right) ^{D_\mathrm{f}-1}. \label{Eq: HRR_function_scaleR}
\end{align}
This formulation highlights that the heat release rate strongly depends on the fractal dimension of the flame contours. A higher fractal dimension value indicates a higher space-filling nature of the flame contours, resulting in enhanced heat release rate. 


Now, consider the evolution of flame contours during a state of sustained periodic acoustic pressure oscillations as shown in Fig.~\ref{fig: flameDynamics_spiky}a-I to Fig.~\ref{fig: flameDynamics_spiky}a-VI. The time series of heat release rate oscillations overlaid with the corresponding fractal dimension values of the flame contours is presented in Fig.~\ref{fig: flameDynamics_spiky}b. The fractal dimension is calculated using the box-counting method (refer Appendix \ref{app_B} for more details). Referring to Fig.~\ref{fig: flameDynamics_spiky}a-I, the flame contour extends along the ${x}$ direction, the longitudinal axis of the combustion chamber, and is characterized by the presence of small-scale wrinkles. The corresponding fractal dimension value of the flame contour is $D_\mathrm{f}=1.39$. During the instance shown in Fig.~\ref{fig: flameDynamics_spiky}a-II, the flame contour rolls up along the ${y}$ direction, and large-scale wrinkles begin to appear on the flame contour, and the fractal dimension increases to 1.45. Subsequently, a large-scale structure characterized by multiple scales, from small to large, emerges in the spatial domain with a fractal dimension value of 1.56 (Fig.~\ref{fig: flameDynamics_spiky}a-III). Followed by this, the large-scale wrinkles begin to disappear and the value of the fractal dimension decreases ($D_{\mathrm{f}}=1.53$ corresponding to Fig.~\ref{fig: flameDynamics_spiky}a-IV and $D_{\mathrm{f}}=1.39$ corresponding to Fig.~\ref{fig: flameDynamics_spiky}a-V). Finally, the flame contour extends along the $x$ direction and is characterized by the presence of small-scale wrinkles with a fractal dimension value of 1.39 (Fig.~\ref{fig: flameDynamics_spiky}a-VI).  


The multiple scales of wrinkles associated with the emergent large-scale structure increase the space-filling of flame contours, leading to an intense heat release rate.  This large-scale structure emerges periodically for each acoustic cycle (Fig.~\ref{fig: CH6 flameDynamics_physical} and Fig.~\ref{fig: flameDynamics_spiky}). This leads to a periodic oscillation of the fractal dimension of flame contours (Fig.~\ref{fig: flameDynamics_spiky}b). 

Referring to Fig.~\ref{fig: flameDynamics_spiky}b, we observe that the heat release rate oscillates with an average phase lag of 57 degrees relative to the fractal dimension of the flame contours. This observed phase difference ($\phi_1$) is due to the time delay associated with turbulent mixing and chemical reactions, which is evident in Fig.~\ref{fig: CH6 flameDynamics_physical}a and explained as follows. Corresponding to the instant at which the large-scale structure appears (Fig.~\ref{fig: CH6 flameDynamics_physical}a-III), only localized patches of heat release rate were observed in the spatial domain (Fig.~\ref{fig: CH6 flameDynamics_physical}b-III). Followed by the emergence of the large-scale structure (Fig.~\ref{fig: CH6 flameDynamics_physical}a-III), after a certain time delay, an intense heat release rate is observed in the spatial domain (Fig.~\ref{fig: CH6 flameDynamics_physical}b-IV).

Referring to Fig.~\ref{fig: CH6 flameDynamics_physical}c, the peak values of the heat release rate and the acoustic pressure oscillation nearly coincide. The average phase difference between the heat release rate and the acoustic pressure oscillation is 8 degrees, a value close to zero. 

Based on the experimental observation of periodic oscillation of the fractal dimension of the flame contours in synchrony with the acoustic pressure,  
\begin{equation}
    D_\mathrm{f}(t)= \bar{D}_\mathrm{f}+A~\mathrm{sin}(\omega t), \label{Eq: FD dynamics}
\end{equation}
provides the simplest representation of the temporal dynamics of the space-filling characteristics of the flame contours. Here, time is chosen such that the initial phase of the fractal dimension is zero. $\bar{D}_\mathrm{f}=1.45$ represents the mean value of the fractal dimension. $\omega=2\pi {f}$, where $f$ is the frequency of the oscillations of the fractal dimension. $A$ represents the extent of modulations in the fractal dimension of the flame contour. Referring to Fig.~\ref{fig: flameDynamics_spiky}b, $D_\mathrm{f}$ varies between 1.35 and 1.55, implying, $A \approx 0.2$.

\begin{figure}[h!]
    \centering
    \includegraphics[width=0.95\linewidth]{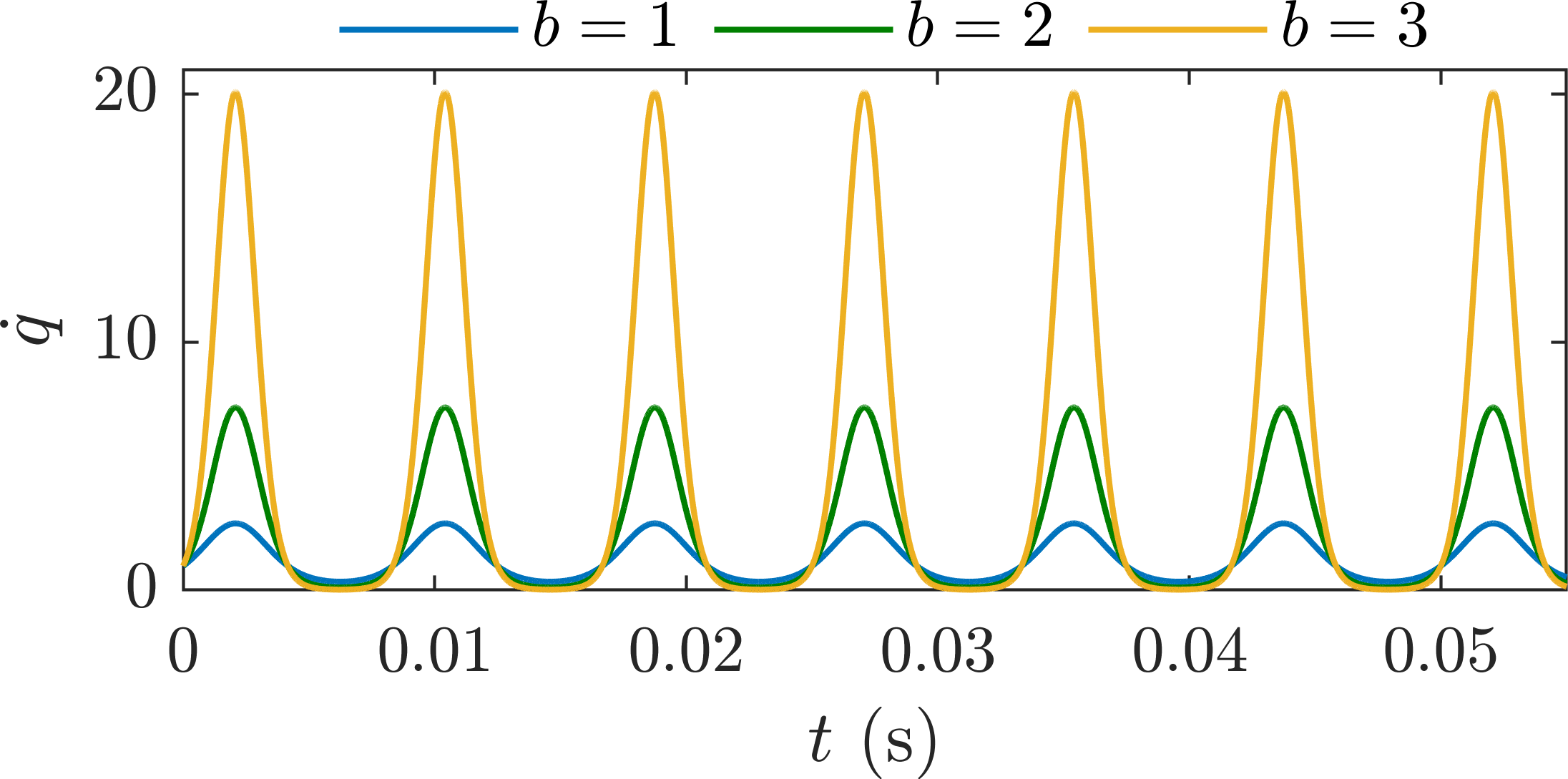}    
    \caption{The variation of  $\dot{q}=ae^{b~\mathrm{sin}(\omega t+\phi_1)}$ for different values of $b$, by considering $\omega=2\pi \times 120$, $a=1$ and $\phi_1=0$.}
    \label{fig: spiky_oscillation_hypothesis}
\end{figure}

Considering Eq.~\eqref{Eq: HRR_function_scaleR}, in which the heat release rate is expressed as a function of the fractal dimension, and substituting the expression for the fractal dimension shown in Eq.~\eqref{Eq: FD dynamics}, we obtain, 
\begin{align}
    \dot{q} &\sim \left(  \frac{L_{\mathrm{o}}}{L_\mathrm{i}}\right)^{\bar{D}_\mathrm{f}-1} \left(  \frac{L_{\mathrm{o}}}{L_\mathrm{i}}\right) ^{A~\mathrm{sin(\omega t)}}. \label{Eq: HRR_function_Df}
\end{align}
We focus on the dynamic evolution of the fractal topology and the scaling exponent of the turbulent flame. Therefore, we rewrite Eq.~\eqref{Eq: HRR_function_Df} by changing its base, from ${L_{\mathrm{o}}}/{L_\mathrm{i}}$ to $e$, grouping all the system parameters, as



\begin{equation}
    \dot{q}(t) \sim a~e^{b~\mathrm{sin}(\omega t + \phi_1)}.
    \label{Eq: spiky_oscillation_hypothesis}
\end{equation}
Here, $a={(L_{\mathrm{o}}/L_{\mathrm{i}})^{\bar{D}_\mathrm{f}-1}}$ and $b=A~\mathrm{log}(L_{\mathrm{o}}/L_{\mathrm{i}})$. The variation of $\dot{q}$ obtained from Eq.~\eqref{Eq: spiky_oscillation_hypothesis} for different values of $b$ is shown in Fig.~\ref{fig: spiky_oscillation_hypothesis}. We observe that these oscillations are reminiscent of the spiky oscillations of heat release rate observed in experiments. For larger values of $b$, the steepness of the spiky oscillation increases. {A larger value of $b$ signifies that the extent of the modulation of the space-filling nature of the flame surface is higher,  therefore steeper the variation in the amplitude of  $\dot{q}$.}

In summary, the periodic emergence of large-scale structure leads to periodic oscillation in the fractal dimension value of the flame contours. Based on this experimental observation, the spiky oscillations in the heat release rate oscillations can be represented as $\dot{q}(t)\sim a~e^{b~\mathrm{sin}(\omega t)}$. 

Pawar \textit{et al.}~\cite{pawar2017thermoacoustic} proposed a mathematical expression, 
\begin{equation}
    \dot{Q}^\prime(t)=[a_1+ \Gamma_1(t)]\times \mathrm{sin}[\omega t + a_2 \mathrm{sin}(\omega t \pm \phi) + \Gamma_2(t)], \label{Eq: spiky_samadhan}    
\end{equation}
using curve fitting to capture the periodic spiky variation in the heat release rate oscillations during the dynamical state of generalized synchronization. Here, $a_1$ and $a_2$ are constants, $\omega$ is the angular frequency of the signal, $\phi$ is the initial phase of the heat release rate oscillations, and $\Gamma_1$ and $\Gamma_2$ are the Gaussian white noise. However, the proposed expression lacked a physical basis.

Next, we examine the spiky heat release rate oscillations during the state of generalized synchronization by considering the mathematical expression - Eq.~\eqref{Eq: spiky_oscillation_hypothesis}, we arrived at by considering the dynamics of the space-filling nature of the flame contours. 



\section{Spiky oscillations of heat release rate during the state of Generalized synchronization}

We begin this section by discussing the synchronous dynamics between the acoustic pressure and the heat release rate oscillations corresponding to the time series shown in Fig.~\ref{fig: experimental setup}b. We use two measures for characterizing their synchronization behavior; phase-locking value and recurrence characteristics.

Phase-locking value (PLV) is a quantitative measure of synchronization between two interacting subsystems \cite{lachaux1999measuring, mormann2000mean}. PLV between the acoustic pressure and the heat release rate oscillations is defined as, 
\begin{equation}
    \mathrm{PLV}_{{p^\prime \dot{q}^\prime}}=\frac{1}{N}  \left| \sum_{t=1}^{N} e^{i \Delta \phi_{p^\prime \dot{q}^\prime}(t)}\right|.
\end{equation}
Here, $\Delta \phi_{p^\prime \dot{q}^\prime}(t)=\phi_{p^\prime}(t)-\phi_{\dot{q}^\prime}(t)$ is the instantaneous relative phase between the acoustic pressure and heat release rate oscillations and $N$ is the total number of points in the time series. PLV values range from 0 to 1; a low PLV value close to 0 indicates desynchronized dynamics, whereas a value close to 1 indicates strong synchronization between the interacting subsystems. PLV value obtained for $p^\prime$ and $\dot{q}^\prime$ shown in Fig.\ref{fig: experimental setup}b is 0.97, indicating strong synchronized dynamics between $p^\prime$ and $\dot{q}^\prime$.

Recurrence is a fundamental property of a deterministic dynamical system \citep{eckmann1995recurrence}. The probability of recurrence ($\mathcal{P}$) measures how frequently a trajectory revisits the neighborhood of a given point in phase space after a time lag $\tau$ \citep{romano2005detection}. A correlation between the recurrence probabilities of two interacting systems signifies synchronized dynamics between them \citep{lakshmanan2011dynamics}. During the occurrence of generalized synchronization, a functional relation between the dynamics of the interacting subsystems implies a near identical recurrence of their trajectory in the phase space \citep{romano2005detection, lakshmanan2011dynamics}.

\begin{figure}[h!]
    \centering
    \includegraphics[width=0.95\linewidth]{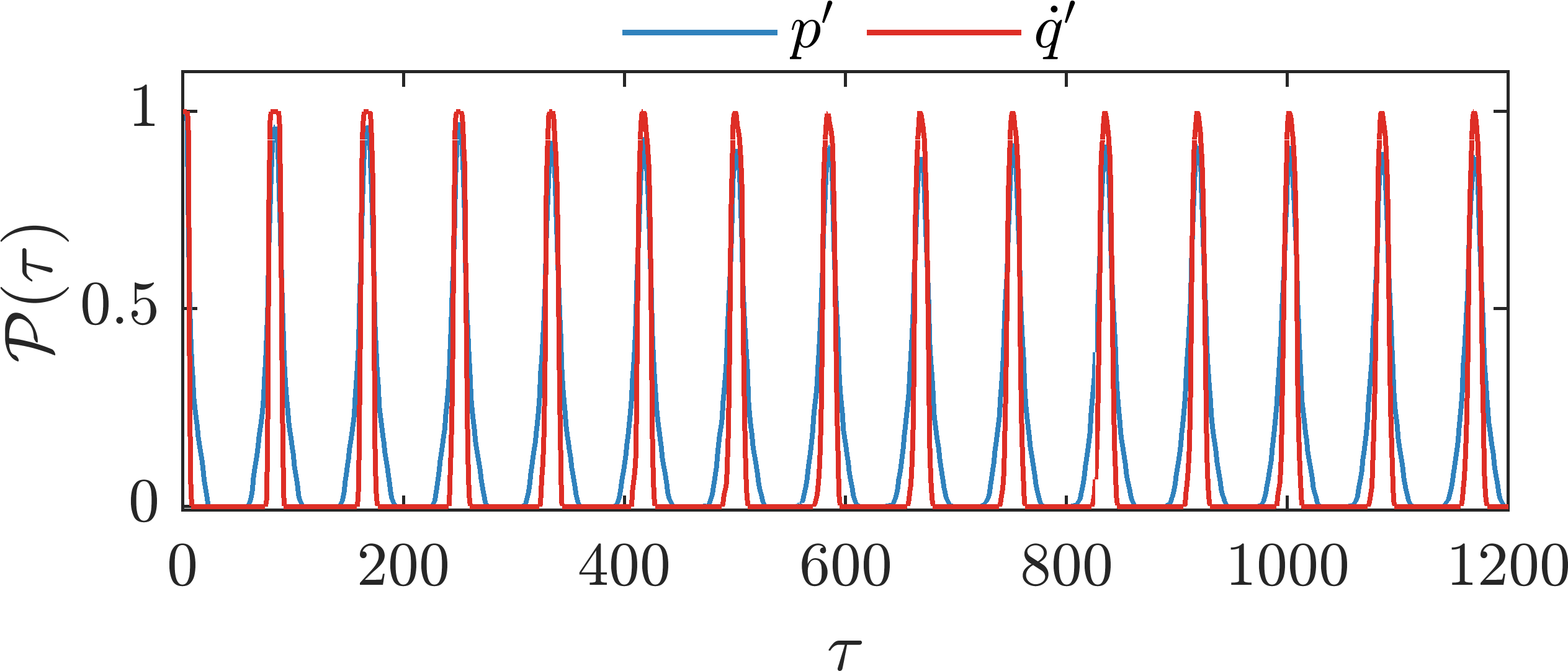}    
    \caption{Probability of recurrence ($\mathcal{P}(\tau)$) for the time series of $p^\prime(t)$ and $\dot{q}^\prime(t)$, shown in Fig.~\ref{fig: experimental setup}b, with time lag $\tau$. The strong correspondence in the recurrence of $p^\prime$ and $\dot{q}^\prime$ is an indication of generalised synchronization between $p^\prime$ and $\dot{q}^\prime$.}
    \label{fig: reccurence_Probability}
\end{figure}


The recurrence probabilities of $p^\prime$ and $\dot{q}^\prime$ with time lag $\tau$ are shown in Fig.~\ref{fig: reccurence_Probability}. The peak values of $\mathcal{P}$ for $p^\prime$ and $\dot{q}^\prime$ are close to 1. Moreover, the peak values of $\mathcal{P}$ for $p^\prime$ coincide with those of the peak values of $\dot{q}^\prime$. 

This strong correspondence in the recurrence of $p^\prime$ and $\dot{q}^\prime$ and a PLV value close to 1 indicates that the coupled dynamics between $p^\prime$ and $\dot{q}^\prime$ correspond a state of generalized synchronization. Next, we look at the spiky variation in the amplitude of the  heat release rate oscillations corresponding to this dynamical state of generlaized synchronization. 


\subsection{Amplitude variation of the heat release can be approximated as $\mathbf{e^{\mathrm{\mathbf{sin(\omega t)}}}}$ \label{CH6: Section-Amplitude variation}}

\begin{figure*}
    \centering
    \includegraphics[width=0.9\linewidth]{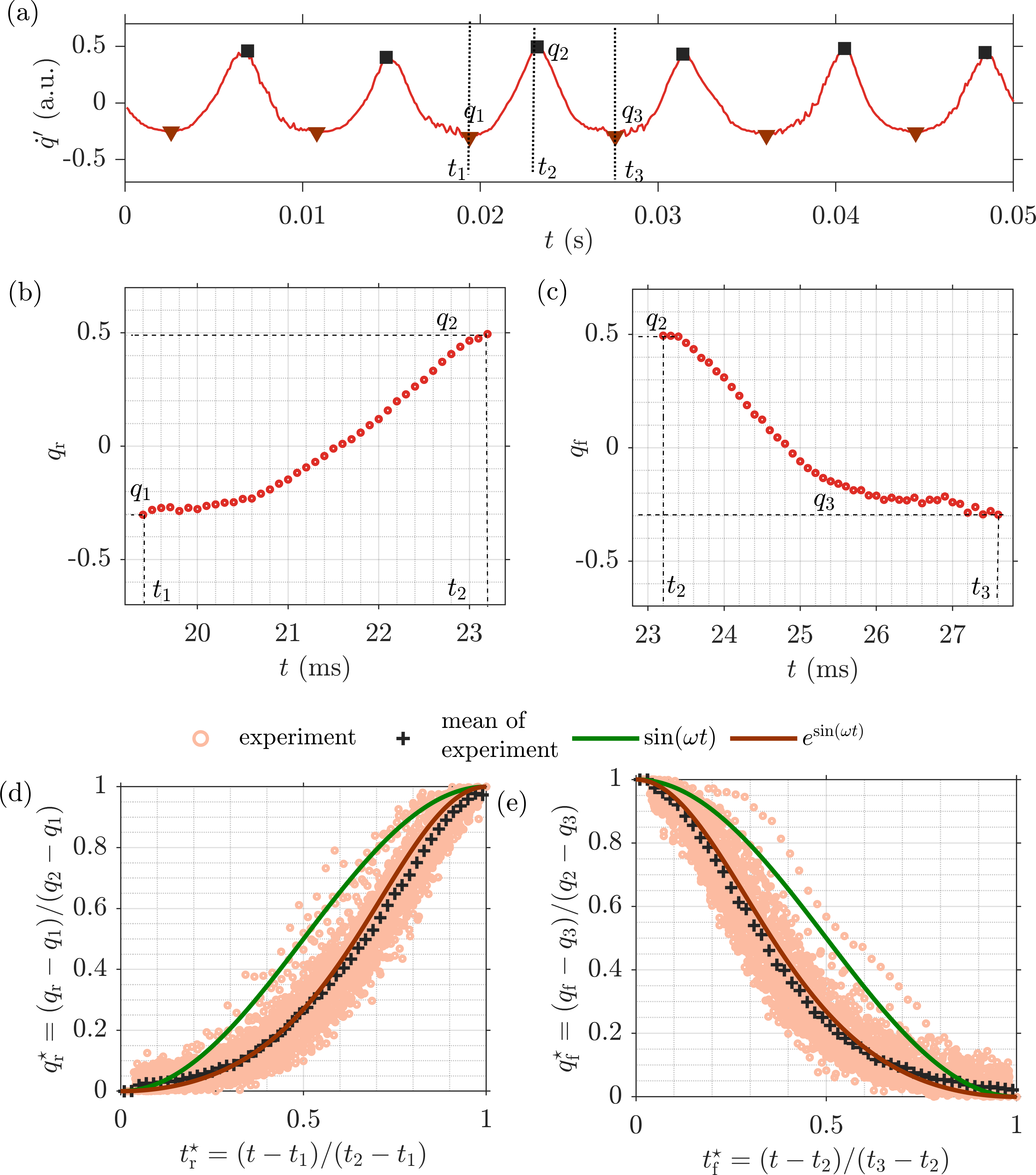}    
    \caption{\textbf{Statistical analysis reveals that the variation of the heat release rate is approximated as $e^{\mathrm{sin(}\omega t)}$.} (a) The time series of heat release rate oscillation with the local maxima ($\blacksquare$) and minima ({\color{brown}\rotatebox[origin=c]{180}{$\blacktriangle$}}) is labelled. The epoch where the variation in the amplitude is from a local minimum to a subsequent local maximum (denoted with $q_{\mathrm{r}}$, point $q_1$ to $q_2$) is referred to as rising epoch ($t_1$ to $t_2$), similarly, the epoch where the variation is from a local maximum to a subsequent local minimum (denoted with $q_{\mathrm{f}}$, point $q_2$ to $q_3$) is referred to as falling epoch ($t_2$ to $t_3$). (b, c) The variation of $q_\mathrm{r}$ and $q_\mathrm{f}$ with respect to time. (d, e) The scaled variations in the amplitude of the heat release rate ($q^\star_\mathrm{r}$, $q^\star_\mathrm{f}$) with respect to the rescaled time ($t^\star_\mathrm{r}$, $t^\star_\mathrm{f}$), respectively, for 100 rising and falling epochs. It is evident that a functional form of $e^{\mathrm{sin}(\omega t)}$ approximates the amplitude variation of the heat release rate.}
    \label{fig: HRR_VARIATION}
\end{figure*}

Figure \ref{fig: HRR_VARIATION}a shows the variation in the amplitude of the heat release rate oscillations during the state of generalized synchronization, with local maxima and local minima indicated. The interval during which the heat release rate increases from a local minimum to the subsequent local maximum is referred to as the rising epoch. For example, during the interval from $t_1$ to $t_2$, the amplitude of $\dot{q}^\prime$ increases from $q_1$ to $q_2$ (Fig.~\ref{fig: HRR_VARIATION}a). In a similar manner, the epoch, while the heat release rate decreases from a local maximum to the subsequent local minimum, is referred to as a falling epoch (for example, from $t_2$ to $t_3$ in Fig.~\ref{fig: HRR_VARIATION}a). Zoomed-in views of the variation of the amplitude during the rising ($q_{\mathrm{r}}$) and the falling ($q_{\mathrm{f}}$) epoch are shown in Fig.~\ref{fig: HRR_VARIATION}b,c, respectively. In order to understand the statistical trend in the variation of the amplitude of heat release rate, we normalize the amplitude and time corresponding to the rising and falling epochs, as follows, 
\begin{align}
q^\star_{\mathrm{r}} &= \frac{q_\mathrm{r} - q_1}{q_2 - q_1}, \quad
t^\star_{\mathrm{r}} = \frac{t - t_1}{t_2 - t_1} \label{eq:rstar}
\end{align}
\begin{center}
    \&
\end{center}
\begin{align}
q^\star_{\mathrm{f}} &= \frac{q_\mathrm{f} - q_3}{q_2 - q_3}, \quad
t^\star_{\mathrm{f}} = \frac{t - t_2}{t_3 - t_2} \label{eq:fstar}
\end{align}
Here, $q^\star$ and $t^\star$ are the normalized amplitude and time. The subscripts  $\mathrm{r}$ and $\mathrm{f}$ correspond to the rising and falling epochs. The normalized amplitudes of the heat release rate oscillations,  $q^\star_{\mathrm{r}}$ and $q^\star_{\mathrm{f}}$ for 100 rising and falling epochs are shown in Figs.~\ref{fig: HRR_VARIATION}d,e, respectively. 


Next, we compare the variations of the normalized variables, $q^\star$ and $t^\star$, obtained from the experimental data with  $q^\star$ and $t^\star$  of the functions $\mathrm{sin}(\omega t)$ and $e^{\mathrm{sin}(\omega t)}$ (Figs.~\ref{fig: HRR_VARIATION}d,e). We have considered the initial phase of these functions to be zero, as the normalized variables remain the same regardless of a phase difference.

We find that the variation in the amplitude of the heat release rate deviates significantly from a sinusoidal form (Fig.~\ref{fig: HRR_VARIATION}d,e). As discussed in Section \ref{section: physical_mechanism_SpikyHRR}, we have deduced a functional form of $\dot{q} \sim e^{b~\mathrm{sin}(\omega t)}$ for the spiky heat release rate oscillations, based on the experimental observation of the periodic oscillation of the fractal dimension of flame contours. Referring to Fig.~\ref{fig: HRR_VARIATION}d,e, we find that the variation in the amplitude of heat release rate is in close agreement with a functional form of,
\begin{equation}
     \dot{q} \sim e^{\mathrm{sin}(\omega t)},
     \label{Eq: HRR_from_statistics}
\end{equation}
with the value of the coefficient, $b \approx 1$, during the state of generalized synchronization. 


Even with the underlying increased complexities of the flame contours, the periodic spiky heat release rate oscillations can be represented in terms of a simple mathematical expression. This observed mathematical expression is consistent across different experimental realizations (refer Appendix \ref{app_C} for more details). In the following section, we discuss the emergent functional relation between the acoustic pressure and the heat release rate oscillations during the state of generalized synchronization. 

\subsection{Synchronization manifold between the acoustic pressure and the heat release rate oscillations}

During the state of generalized synchronization, a functional relationship exists between the dynamics of the coupled oscillators \citep{rulkov1995generalized}. For example, if $X(t) \in \mathbb{R}^m$, and $Y(t) \in \mathbb{R}^n$ represent the dynamics of two oscillators, then a functional relationship, $Y(t)=\Phi(X(t))$ exists between the oscillators.

\begin{figure*}[ht]
    \centering
    \includegraphics[width=0.9\linewidth]{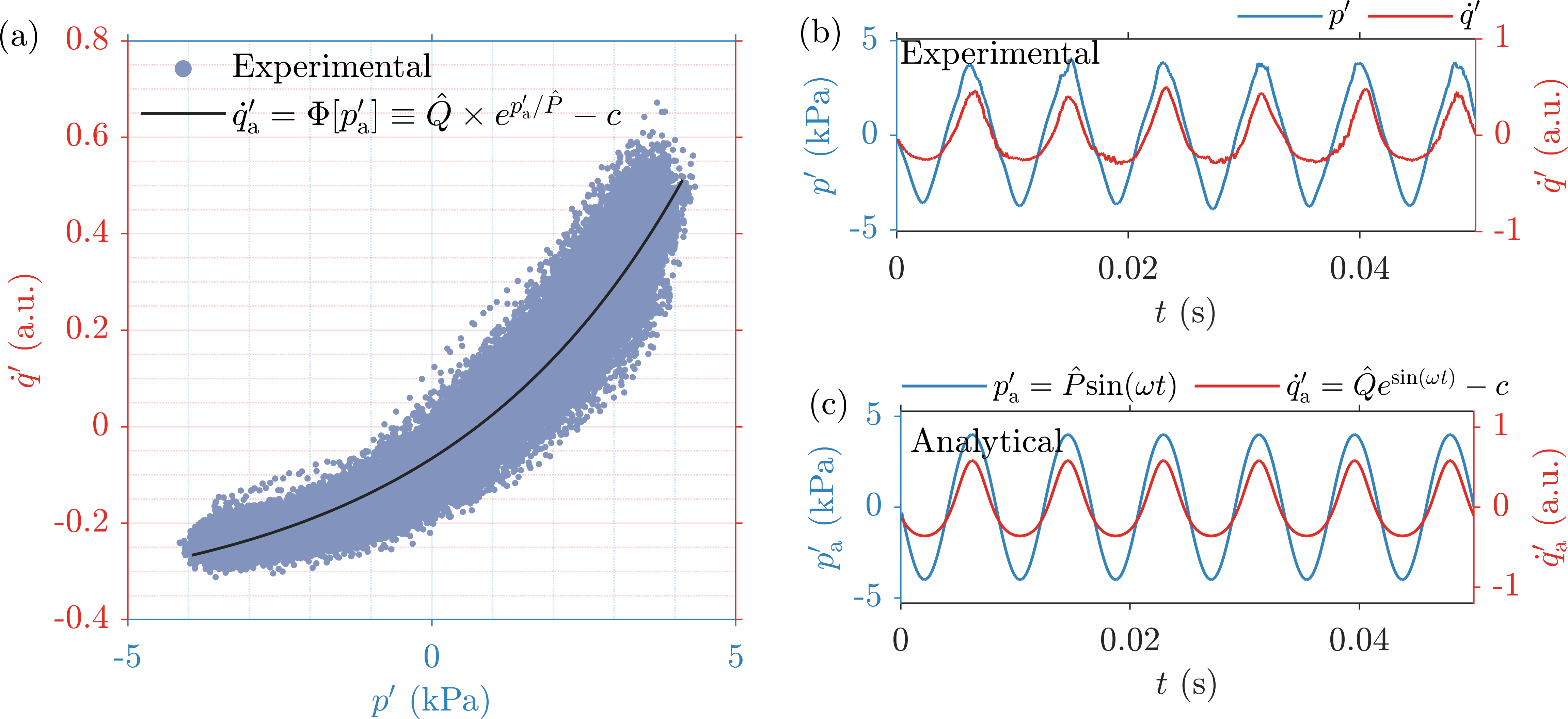}    
    \caption{ \textbf{Synchronization manifold during the state of generalized synchronization between the acoustic pressure and the heat release rate oscillations.} (a) The heat release rate variation with respect to the acoustic pressure.  The experimental data points are scattered around a subspace, which has a functional form, $\dot{q}^\prime_{\mathrm{a}}=\Phi[p^\prime_\mathrm{a}]$, with $\Phi=\hat{Q} \times e^{p^\prime/\hat{P}}-c$ is indicated with a solid line. The scattered experimental points around the functional manifold, $\Phi$, suggest that the system is perturbed continuously due to the underlying turbulent fluctuations. (b) The time series of the acoustic pressure (blue) and the heat release rate oscillations (red) obtained from the experiments ($p^\prime$ and $\dot{q}^\prime$) (c) and the time series corresponding to the analytic expressions $p^\prime_\mathrm{a}= \hat{P}~\mathrm{sin}(\omega t)$ and $\dot{q}^\prime_\mathrm{a}= \hat{Q} e^{\mathrm{sin}(\omega t)}-c$. We find that the variation of the heat release rate obtained from the analytic expression is in close agreement with the experiments.}
    \label{fig: synch_manifold}
\end{figure*}


In our turbulent reactive flow system, an approximate functional relationship between the acoustic pressure ($p^\prime$) and the unsteady heat release rate ($\dot{q}^\prime$) during the state of generalized synchronization can be inferred based on the insights from experimental observations.  Referring to Fig.~\ref{fig: experimental setup}b, the acoustic pressure oscillations can be represented as a sinusoidal function \citep{pawar2017thermoacoustic}. The spiky variation in the heat release rate, based on the physical mechanism and the statistics of the variation of the amplitude of the heat release rate, can be approximated by an expression of the form given in Eq.~\eqref{Eq: HRR_from_statistics}. Therefore, the acoustic pressure ($p^\prime_{\mathrm{a}}$) and the heat release rate ($\dot{q}^\prime_{\mathrm{a}}$) can be expressed as,     
\begin{align}
    p^\prime_{a} &=  (\hat{P}+\Gamma_1(t)) \mathrm{sin} (\omega t), \label{Eq: p_analytical} \\
    \dot{q}^\prime_{a} &=  (\hat{Q}+\Gamma_2(t)) e^{\mathrm{sin}(\omega t + \phi_2)}-c. \label{Eq: q_analytical}    
\end{align}
Here, $\hat{P}$ and $\hat{Q}$ are the time-dependent amplitudes of the oscillations, and  $c=<(\hat{Q}+\Gamma_2(t)) e^{\mathrm{sin}(\omega t)}>$, where $<>$ represents the mean value. $c$ is subtracted in Eq.~\eqref{Eq: q_analytical} to obtain the mean-subtracted fluctuations. From the experimental data, the following values are obtained: the amplitude of acoustic pressure oscillations, $\hat{P}=4.5$ kPa; amplitude of heat release rate oscillations, $\hat{Q}=0.40$ a.u.; and $c=0.46$ a.u. $\phi_2$ is the phase difference between the acoustic pressure and the heat release rate oscillations, whose value is close to 0 degrees. $\Gamma_1$ and $\Gamma_2$ are the white Gaussian noise added to the amplitudes of the acoustic pressure and the heat release rate oscillations. The white Gaussian noise is added to introduce the fluctuations arising from the underlying hydrodynamic turbulent flow. 

Now consider the variation of the heat release rate with respect to the acoustic pressure oscillations obtained from the experiment (Fig.~\ref{fig: synch_manifold}a). It is evident that the acoustic pressure and the heat release rate variations are correlated, and the data points are restricted to a subspace in the $p^\prime-\dot{q}^\prime$ plane during the state of generalized synchronization (Fig.~\ref{fig: synch_manifold}a). The spread of experimental data is due to the continuous perturbations to the synchronized state arising from the underlying incoherent part of the turbulent flow. 

From Eqs.~\eqref{Eq: p_analytical}-\eqref{Eq: q_analytical} and based on the experimental observation as shown in Fig.~\ref{fig: synch_manifold}a, the emergent functional relation between the subsystems, representing the synchronization manifold, can be obtained. We consider $\phi_2=0$ in Eq.~\eqref{Eq: q_analytical} since the phase difference between the acoustic pressure and the heat release rate is close to zero.  Omitting the noisy perturbations to the synchronized state, i.e., $\Gamma_1=0$ and $\Gamma_2 =0$, and by assuming $p^\prime_{a}/\hat{P}=\mathrm{sin}(\omega t)$, it is straightforward to obtain the functional relation, $\Phi$, from  Eq.~\eqref{Eq: p_analytical} and Eq.~\eqref{Eq: q_analytical}.  The expression for $\Phi$ is as follows,
\begin{equation}
    \dot{q}^\prime_{\mathrm{a}}=\Phi[p^\prime_{\mathrm{a}}]\equiv \hat{Q} \times e^{p^\prime_{\mathrm{a}}/\hat{P}}-c. \label{Eq: manifold}
\end{equation}
The functional relation for $\Phi$ in Eq.~\eqref{Eq: manifold} is plotted along with the experimental data (Fig.~\ref{fig: synch_manifold}a). The time series of acoustic pressure and the time series of heat release rate obtained from the analytical expressions, shown in Eqs.~\eqref{Eq: p_analytical}-\eqref{Eq: q_analytical}, match very well with the experimental data shown in Figs.~\ref{fig: synch_manifold}b,c, respectively.

\section{Conclusion and discussion}
In a turbulent reactive flow system, we observe that the heat release rate exhibits sustained periodic spiky oscillations in synchrony with the sinusoidal acoustic pressure oscillations. We have studied the physical processes, i.e., the dynamics of the space-filling nature of the flame, associated with these spiky oscillations. 

The spiky oscillations arise due to the slow and fast time scales associated with the underlying physical processes. Vortices shed from the inlet step of the combustor entrain the reactants and convect downstream, and collectively interact to form a large-scale coherent structure. The emergence of this large-scale structure is associated with the slow convective timescale. Subsequently, the enhanced space-filling nature of the flame contours, due to the large-scale structure, leads to a sudden increase in the heat release rate on a fast timescale. The large-scale structure emerges periodically with a time period corresponding to the acoustic pressure cycle, analogous to the refractory time period, characteristic of oscillations in excitable media. The periodic emergence of the large-scale structure leads to a periodic oscillation in the fractal dimension of the flame contours, resulting in periodic spiky heat release rate oscillations. We find that our system exhibits the features of spiky oscillations that are ubiquitous in excitable media. Our analysis provides the first evidence that the periodic dynamics in the fractal topology of a propagating wave front can lead to spiky oscillations.

Based on the periodic oscillation in the fractal dimension of the topology of the flame surface and the statistical analysis of the variation of the amplitude of the heat release rate oscillations, we find that the spiky oscillations can be approximated by a functional form of $e^{\sin(\omega t)}$ during the state of generalized synchronization between the heat release rate and the acoustic pressure. Further, we unravel the emergent functional relation representing the synchronized dynamics between the acoustic pressure and the heat release rate oscillations during the state of generalized synchronization.  It is intriguing that the dynamics of such a highly nonlinear, far from equilibrium turbulent reactive flow system can be represented by simple mathematical equations.

\begin{acknowledgments}

We acknowledge Mrs. Shruti Tandon for the fruitful discussions and feedback. 
We acknowledge Dr. Nitin Babu George and Mr. Midhun P. R. for providing the experimental data.  Sivakumar Sudarsanan is grateful to the Ministry of Education, Govt of India for the Half-Time Research Assistantship (HTRA). R.I. Sujith expresses his gratitude to the Department of Science and Technology and Ministry of Human Resource Development, Government of India, for providing financial support for our research work under Grant No.
SP22231222CPETWOCTSHOC (Institute of Eminence grant). 
\end{acknowledgments}

\appendix

\section{Method for identifying the flame contours\label{app_A}}

Planar Mie scattering images of the flow field seeded with TiO\textsubscript{2} particles were recorded by illuminating the flow field with a laser sheet. Scattered light from the TiO\textsubscript{2} particles was captured using a high-speed camera.  Across the flame, two distinct regions with different particle number densities are observed. The high-density regions represent the reactant side, whereas the low-density regions represent the product side.  Thus, the interface between the regions of high and low  TiO\textsubscript{2} particle density in the reactive flow indicates the topology of the flame \citep{stella2001three}. 

\begin{figure}
\centering
\includegraphics[width=\linewidth]{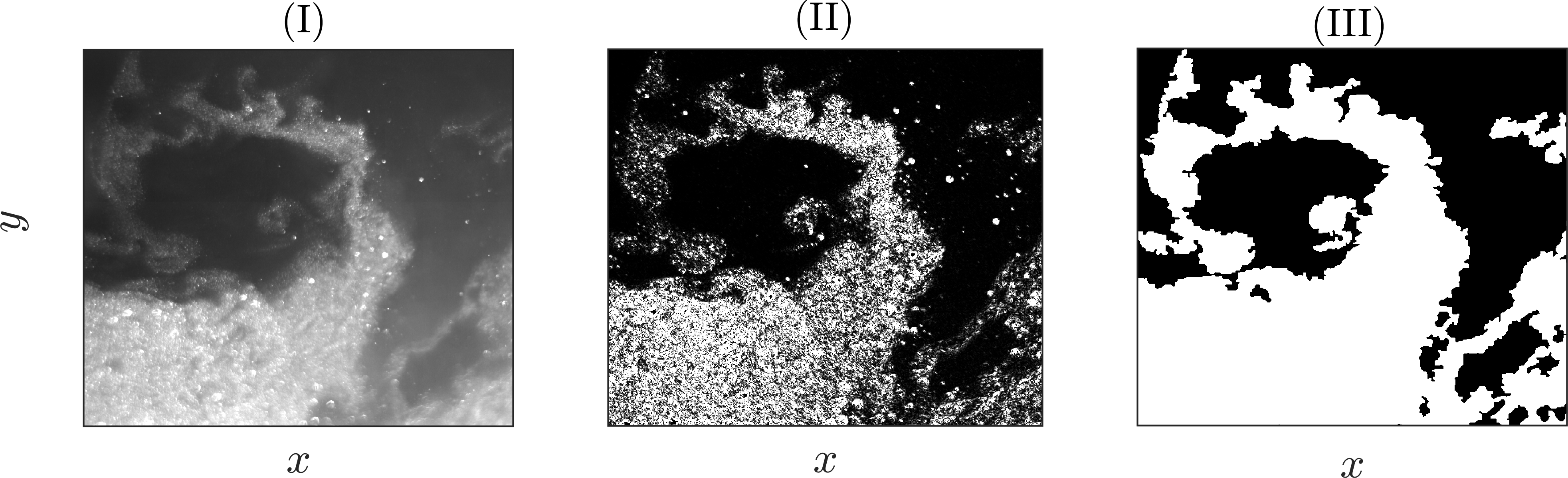}
\caption{(I) TiO\textsubscript{2} image ($\mathbf{M}$) recorded at an instant during $Re=2.78 \times 10^4$ and $Re=3.57 \times 10^4$. (II) The measure $\mathbf{G}$ obtained from $\mathbf{M}$ corresponding to (I). The binarised image after setting a threshold value to $\mathbf{G}$. Black represents the regions with low particle density, whereas white colour represents the regions with high particle density.}
\label{Fig: appendix_flameContour}
\end{figure}

The TiO\textsubscript{2} image is a two-dimensional matrix (Fig.~\ref{Fig: appendix_flameContour}I), represented by $\mathbf{M}$, with its entries representing the intensity values of the scattered light from the flow field.  Our strategy for identifying the flame contour is to detect the TiO\textsubscript{2} particles and then use a morphological operation to differentiate the regions of high and low particle number density. The steps for obtaining the flame contour are described as follows.

First, we construct a measure $\mathbf{G}$ by combining the intensity field $\mathbf{M}$ with its first- and second-order spatial derivatives to enhance the particle characteristics. The change in the intensity value across a particle is captured by the first-order spatial derivative of $\mathbf{M}$, denoted as $\mathbf{M}^\prime$. We compute the spatial derivative (along x and y directions) by finding the convolution of $\mathbf{M}$ with the Sobel kernel matrix ($\mathbf{K}$) \citep{sobel1970camera}, expressed as follows 

\begin{equation}
\begin{aligned}
    \mathbf{M}_{x}^\prime &= (\mathbf{M} * \mathbf{K})_{i,j} = \sum_{m}\sum_{n} \mathbf{M}_{i-m,j-n}\mathbf{K}_{m,n}.
\end{aligned}
\end{equation}

Here $*$ denotes the convolution operation, and the kernal matrix $\mathbf{K}$ is defined as \[
\mathbf{K} = 
\begin{bmatrix}
    -1 & 0 & 1 \\
    -2 & 0 & 2 \\
    -1 & 0 & 1
\end{bmatrix}.
\]
Similarly, we compute the gradient along the y direction, and then compute $\mathbf{M^\prime}$ as follows,  
\begin{equation}
\begin{aligned}
    \mathbf{M}_{y}^\prime &= (\mathbf{M} * \mathbf{K}^{\mathsf{T}})_{i,j} , \\
    \mathbf{M}^\prime &= \mathbf{M}_{x}^\prime+\mathbf{M}_{y}^\prime .
\end{aligned}
\end{equation}
Here, $\mathbf{K}^{\mathsf{T}}$ is the transpose of $\mathbf{K}$. The Sobel kernel is widely used in image processing for edge detection, with applications across different fields, for example, medical imaging \citep{litjens2017survey}, and remote sensing \citep{zhang2016deep}.

Next, the second-order derivatives are obtained by 

\begin{equation}
\begin{aligned}
    \mathbf{M}_{x}^{\prime\prime} &= (\mathbf{M}_{x}^\prime * \mathbf{K})_{i,j} \\
    \mathbf{M}_{y}^{\prime\prime} &= (\mathbf{M}_{y}^\prime * \mathbf{K}^{\mathsf{T}})_{i,j} \\
    \mathbf{M}^{\prime\prime} &= \mathbf{M}_{x}^{\prime\prime}+\mathbf{M}_{y}^{\prime\prime}.
\end{aligned}
\end{equation}
Second-order derivatives capture the peak gradient magnitude along particle boundaries.  We emphasise all these features to arrive at a measure $\mathbf{G}$, defined as 
\begin{equation}
\begin{aligned}
    \mathbf{G}=\mathbf{M} \circ \mathbf{M}^\prime \circ \mathbf{M}^{\prime \prime},  
\end{aligned}
\end{equation}
where $\circ$ represents the element-wise matrix multiplication operator.

TiO\textsubscript{2} image recorded at an instant corresponding to chaotic and periodic oscillatory state is shown in Fig.~\ref{Fig: appendix_flameContour}a-I, b-I. The obtained measure, $\mathbf{G}$ for these two images are shown in Fig.~\ref{Fig: appendix_flameContour}a-II, b-II. Observing the figures \ref{Fig: appendix_flameContour}a-II, b-II, the regions with high particle density are now distinguishable from the regions with low particle number density. The next step is to employ a thresholding method to $\mathbf{G}$. We select a threshold of $\mathbf{G}_{\mathrm{th}}=<G>+\sigma_{\mathbf{G}}/2$, where $<G>$ and $\sigma_{\mathbf{G}}$ represent the mean and standard deviation of all the elements of $\mathbf{G}$. After the thresholding method, we obtain a binary matrix, and a morphological operation is performed to obtain the regions of high and low particle number density as shown in Fig.~\ref{Fig: appendix_flameContour}a-III, b-III. The interface between these regions is then identified as the flame contour. 

\section{Calculating the fractal dimension of a geometry using box-counting method \label{app_B}}

In this study, the topology of the flame contours arising in the system is analyzed within the framework of fractal geometry. A box-counting algorithm is employed to determine the {fractal dimension} of flame contours in two-dimensional space. In this method, a grid of square boxes with side length ($\epsilon$) is overlaid on the flame contour. The number of boxes $N(\epsilon)$ required to cover the entire flame contour is recorded for a range of box sizes. The scaling behavior of $N(\epsilon)$ with respect to $\epsilon$ is then analyzed, following the relationship, $N(\epsilon) \sim \epsilon^{-D_{\mathrm{f}}}$. A log-log plot of $N(\epsilon)$ against $\epsilon$ is constructed, and the slope of the best-fit line provides an estimate of the fractal dimension $D_{\mathrm{f}}$ for the flame contour.

\begin{figure}[h!]
\centering
\includegraphics[width=\linewidth]{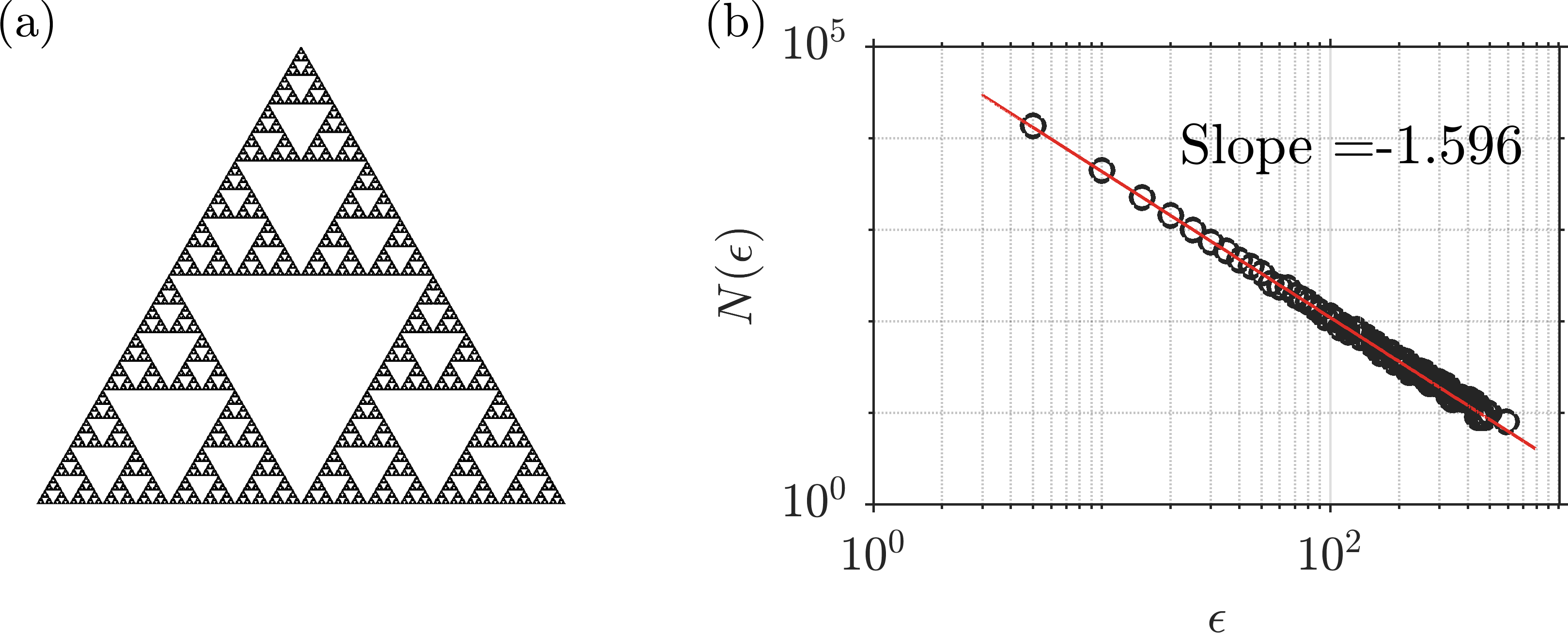}
\caption{(a) The Sierpinski triangle and (b) the box-counting result. We find that the slope of the variation of the number of boxes with respect to box size in the log-log scale is -1.596, which is very close to the theoretical fractal dimension value of the Sierpinski triangle.}
\label{Fig: FD_methodology}
\end{figure}

To validate the box-counting algorithm, we have calculated the fractal dimension of a well-known theoretical fractal, the Sierpinski triangle. The theoretical fractal dimension is log(3)/log(2) $\approx$ 1.585. Using the box-counting algorithm, the obtained value of the fractal dimension is 1.596, with a difference of 0.011.  

\section{Statistics of amplitude variation of heat release rate \label{app_C}}

\begin{figure}[h!]
\centering
\includegraphics[width=\linewidth]{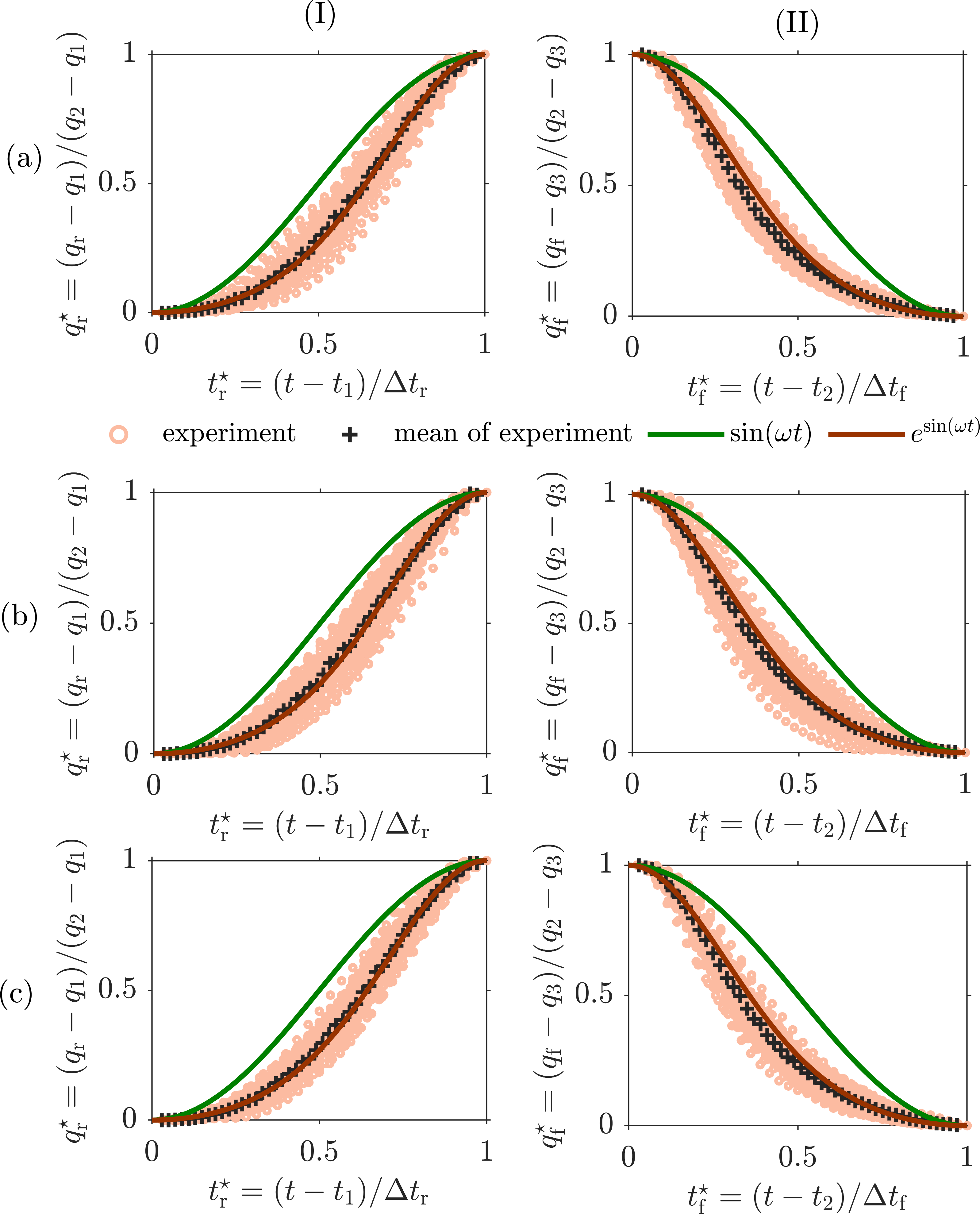}
\caption{(a-c) The variation in the amplitude of rising (column I) and falling (column II) epochs for 100 cycles during the state of generalized synchronization corresponding to three different experimental realizations. The variation of the heat release rate is in close agreement with the mathematical equation, $e^{\mathrm{sin}(\omega t)}$.}
\label{Fig: CH6 Experiment_repeatability}
\end{figure}

Here, we present our results on the statistics of the variation in the amplitude of heat release rate oscillations for three different experimental realizations in a turbulent reactive flow system, performed at a Reynolds number value of $6.53 \times 10^4$ (Figs.~\ref{Fig: CH6 Experiment_repeatability}a-c). These results correspond to the dynamical state of generalized synchronization. The variation in the amplitude for the rising and falling epochs is normalized. We find that the amplitude variation of the heat release rate is in close agreement with a functional form, $e^{\mathrm{sin}(\omega t)}$.


\bibliography{apssamp}

\end{document}